\documentclass[11pt]{article}

\usepackage[margin=0.75in]{geometry}
\usepackage[utf8]{inputenc}
\usepackage[title]{appendix}
\usepackage[sort&compress,square,comma,numbers]{natbib}
\usepackage[version=4]{mhchem}
\usepackage{graphicx,subfig,float}
\usepackage{tabularx}
\usepackage{amsmath,amsthm,amssymb,commath,amsfonts}
\usepackage{booktabs,multirow}
\usepackage[normalem]{ulem}
\usepackage[colorlinks=true,bookmarksopen=true]{hyperref}
\usepackage{algpseudocode}
\usepackage{algorithm}
\usepackage{setspace}
\usepackage[normalem]{ulem}
\usepackage{todonotes}
\usepackage{multirow}

\usepackage{dcolumn}
\usepackage{bm}

\usepackage{siunitx}

\begin{document}

\title {Confinement-mediated phase behavior of hydrocarbon fluids:\\Insights from Monte Carlo simulations\thanks{Approved for public release (INL/JOU-20-57581)}}

\author{
Jiaoyan Li, Qi Rao, Yidong Xia\thanks{Corresponding author: Yidong Xia (\url{yidong.xia@inl.gov})}\\
Energy and Environment Science \& Technology, Idaho National Laboratory\\ 1955 N. Fremont Ave., Idaho Falls, ID 83415-2025, USA\\
\and
Michael P. Hoepfner, Milind Deo\\
Department of Chemical Engineering, The University of Utah\\50 Central Campus Dr, Salt Lake City, UT 84112, USA\\
}

\date{}

\maketitle

\section*{\centering Abstract}

The phase behavior of hydrocarbon fluids confined in porous media has been reported to deviate significantly from that in the bulk environment due to the existence of sub-10nm pores. Though experiments and simulations have measured the bubble/dew points and sorption isotherms of hydrocarbons confined in both natural and synthetic nanopores, the confinement effects in terms of the strength of fluid-pore interactions tuned by surface wettability and chemistry have received comparably less discussion. More importantly, the underlying physics of confinement-induced phenomena remain obfuscated. In this work, we studied the phase behavior and capillary condensation of n-hexane to understand the effects of confinement at the molecular level. To systematically investigate the pore effects, we constructed two types of wall confinements; one is a structureless virtual wall described by the Steele potential and the other one is an all-atom amorphous silica structure with surface modified by hydroxyl groups. Our numerical results demonstrated the importance of fluid-pore interaction, pore size, and pore morphology effects in mediating the pressure-volume-temperature (PVT) properties of hydrocarbons. The most remarkable finding of this work was that the saturation pressure predicted from the van der Waals-type adsorption isothermal loop could be elevated or suppressed relative to the bulk phase, as illustrated in the graphical abstract. As the surface energy (i.e., fluid-pore interaction) decreased, the isothermal vapor pressure increased, indicating a greater preference for the fluid to exist in the vapor state. Sufficient reduction of the fluid-pore interactions could even elevate the vapor pressure above that of the bulk fluid.

\vspace{1em}
\noindent {\it Keywords:} Confinement effects, phase behavior, Gibbs Ensemble Monte Carlo, molecular simulations
\section{Introduction}

To date, fossil fuels are still the primary resource used for energy production in the United States. Conventional and unconventional fossil fuels differ in their geologic locations and accessibility. Conventional fuels are often found in easily accessible reservoirs, yet, unconventional fuels exist in low-permeability shale formations and require advanced extraction technologies, such as hydraulic fracturing. Oil/gas shale rocks are porous media, containing sub- 10nm to over 100nm pores, which significantly influence the thermodynamic behaviors of hydrocarbons, result in a reduction of energy production, and cause uncertainties in production estimations. To enhance oil/gas recovery from shale and provide a more accurate estimation of reservoir production, it is necessary for us to study the phase behavior of hydrocarbons under confinement and obtain a comprehensive understanding of the pore effects.

Several experiments have been conducted to study the thermodynamic properties of hydrocarbons and hydrocarbon mixtures confined in porous media. Controlled porous glasses (CPGs), MCM-41 and SBA-15 are commonly used porous media in experimental conditions to resemble the pore network in shale when investigating the confinement-induced phenomena. Current experimental methods include the differential scanning calorimetry (DSC) method, use of nanochannel chips, and adsorption-desorption measurements. Review papers \cite{BARSOTTI2016344, SALAHSHOOR201889, LIU2019102901} have summarized the experimental work based on the methods as mentioned above and are recommended for interested readers. As \citet{LIU2019102901} discuss, there exist controversies concerning bubble point deviation and critical density shift as a result of confinement. For example, \citet{cho2017bubble} and \citet{pathak2017experimental} measured the bubble point of decane-methane and octane-methane mixtures in 9:1 mole ratio confined in SBA-15 and SBA-16 using DSC and concluded that the bubble point temperature was higher than that of bulk mixtures. \citet{qiu2019simple} measured the dew points of a binary methane/ethane gas mixture confined in SBA-15 with differnet pore diameters using similar technique and found that the dew-pressure curve of the fluid mixtures confined in nanopores was below that of the surrounding bulk fluid. \citet{luo2016effect} observed that the bubble point temperature of octane and decane mixture confined in the CPGs had two states; the bubble point temperature of the mixture attached to the CPGs pore wall was lower than the bulk, while that of the mixture in the middle of the pore was higher than the bulk. Based on this observation, they further proposed a two-state model for the hydrocarbons confined in the cylindrical pore. Recently, \citet{luo2018use} measured the bubble point temperatures of octane-decane binary mixtures using DSC for various loading percentage of fluids, and results showed that the bubble point temperature was lower than that of bulk predicted by Peng-Robinson Equation of State (PR-EOS) \cite{peng1976new}. 

Compared with experimental measurements, theoretical models and molecular simulations have the advantage of readily studying extreme conditions (such as high temperature/pressure and nano-/micro- scale phenomena) that are challenging to be measured by experiments. A large number of theoretical models based on an equation of state (EOS) coupled with capillarity \cite{luo2016confinement, dong2016phase, sandoval2017influence, zhang2019phase} predict that the bubble point pressure of nanoconfined hydrocarbon mixtures is reduced, in which the swelling factor, criticality, bubble point, and pore size were studied numerically. In nanopores, the inhomogeneous molecular distribution is significant and plays a critical role in altering the phase boundary and capillary pressure. Therefore, theoretical models of engineering density-functional theory (DFT) combined with an equation of state have been proposed and developed \cite{li2014phase, jin2016thermodynamic, jin2016phase, liu2018comparison, JIN2018177} to explicitly account for this phenomena. \citet{JIN2018177} used engineering DFT to study the effects of nanopore size (under 20nm) on the bubble/dew point and adsorption hysteresis, and found that dew point of confined fluids approached the bulk saturation point as pore size increases, but bubble point was significantly different from bulk even for very large pores. This study represented the first time that a transitional variation of pore size effect on the pressure-temperature diagram was reported in the literature. By comparing these results to grand canonical Monte Carlo (GCMC) simulations, engineering DFT was proven to be able to predict the vapor-liquid equilibrium reliably \cite{JIN2018177}. The predictions from engineering DFT were also compared against experiments conducted in a nanofluidic channel and showed an underestimation of capillary pressure \cite{zhong2018nanoscale}.

Monte Carlo (MC) molecular simulations have been widely used to support experiments and validate theoretical models for the investigation of fluid phase behavior under confinement \cite{JIN2018177, liu2011monte, siderius2017quasi, lowry2018effect, dantas2019phase, jin2016phase2, bi2019molecular}. The inhomogeneous molecular distribution and the fluid-pore interactions of different absorbents and hydrocarbons are more accurately described at the molecular level by MC methods in contrast to theoretical models. Thus, MC molecular simulations are considered as numerical experiments and present molecular behavior at a high resolution, and more importantly, they are expected to provide scientific insights to understand pore effects. Generally speaking, there are two types of MC simulations widely adopted to model phase behavior in confined geometries, i.e., grand canonical Monte Carlo (GCMC) and Gibbs ensemble Monte Carlo (GEMC). GCMC simulations can be used to construct the hysteresis loop formed by discontinuous adsorption and desorption isotherms, in which sudden jumps correspond to spontaneous condensation and evaporation. In contrast, the GEMC method is based on the construction of a continuous adsorption isotherm in the form of a van der Waals loop, which includes stable, metastable, and unstable equilibrium states. Therefore, it is used to determine the equilibrium transition between vapor-like and liquid-like states from the spinodals.

The GEMC method was originally developed to simulate the phase equilibrium of pure substances and mixture systems in bulk environments \cite{panagiotopoulos1987direct, panagiotopoulos1988phase, panagiotopoulos2002direct, kholmatov2011monte, silva2015gibbs} and then later extended to confined systems \cite{panagiotopoulos1987adsorption}. \citet{neimark2000gauge} further developed the GEMC to gauge-cell GEMC (nvt-GEMC) and proposed that the saturation point divided the thermodynamic integration along the metastable and unstable regions of the isotherm equally according to Maxwell's rule of equal areas \cite{neimark2000gauge, vishnyakov2001studies, vishnyakov2009multicomponent}. The unique feature of the GEMC and gauge-cell GEMC is that two or more simulation boxes are adopted to enable molecular swap between boxes to ensure the equilibrium of chemical potential. In this way, GEMC and gauge-cell GEMC can naturally separate the fluids of vapor and liquid phases or fluids in bulk and confined environments. Using this method, \citet{vishnyakov2001studies} studied the capillary condensation of argon in cylindrical pores of different diameters (1.5-5.5nm) and found good agreement with experimental data for equilibrium transitions of argon in pores wider than 2.2nm. Further, hysteretic adsorption-desorption isotherms were predicted in pores wider than 5nm at 87K. Using the GEMC method in the grand canonical and microcanonical ensembles, \citet{dantas2019phase} investigated the influence of temperature on carbon dioxide adsorption and attempted to predict the transition from reversible capillary condensation to hysteretic adsorption-desorption cycles as observed by experiments with a decrease of temperature \cite{dantas2019phase}. Based on the idea of GEMC and gauge-GEMC/nvt-GEMC, the $\mu$vt-GEMC \cite{jin2016phase2} was proposed and applied to a ternary system (C$_1$/C$_3$/C$_5$) under a 4nm slit pore confinement. The $\mu$vt-GEMC simulation approach was also employed to extend the discussion for single pore system to multiple pore system \cite{jin2017molecular} considering that real shale systems usually have broad pore size distributions. A single-component system, methane, was studied as the working fluid, and the results showed that the small pores were filled before the large ones, which indicated that the liquid first condensed in small pores. Recently, the constant composition expansion experiment was simulated by the npt-GEMC for multi-component hydrocarbon mixtures in a multi-scale porous media \cite{bi2019molecular} and serve as a good reference for the study of complex systems.

Despite numerous studies on the liquid-vapor equilibrium of confined fluids from experiments, theoretical models and molecular simulations, the effects of varying the fluid-wall interactions on the phase behavior are rarely reported. Here, the fluid-wall interaction is referred to in a more general way to account for the surface energy of different pore materials, surface wettability, and chemistry. The mineralogy and surface properties of natural shale systems are more complex than the commonly used porous media CPGs, and motivate in-depth studies to fully explore the effect of fluid-wall interaction on the phase behavior and capillarity of confined hydrocarbons. \citet{srivastava2017pressure} proposed a theoretical model to examine the effect on the tangential pressure of varying the molecular shape, the strength of the fluids-wall interactions for monomer, dimer, and trimer confined in carbon slit-shaped pores, and demonstrated that the fluid-wall interactions have a significant impact on the pressure tensors compared to the molecular shape \cite{srivastava2017pressure}. \citet{lowry2018effect} employed both experimental and modeling techniques to investigate the effects of surface chemistry on the adsorption and thermodynamic behaviors of propane and n-butane, in which the surface chemistry was modified by altering the coverage of carbon \cite{lowry2018effect}. An understanding of the effects of fluid-wall interactions is not only important to provide a high-fidelity prediction for the thermodynamic behavior of hydrocarbons in nature shale but also critical to advance CO$_2$ enhanced shale gas recovery techniques \cite{zhang2020recovery}.

In this paper, we studied how confinement modified the phase behavior and capillarity of hydrocarbon fluids as a function of the strength of the fluid-pore interaction, pore size and morphology. The nvt-GEMC simulation method was used to identify the phase coexistence boundary of both bulk and confined n-hexane (nC$_6$). The methodology developed in this work is generally applicable to all other hydrocarbon fluids. To tune the interaction between the pore wall and the working fluid, we introduced an interface intensity  coefficient in the Steele potential \cite{steele1973physical, siderius2011extension} to vary the surface energy and fluid-pore interaction strength induced from a structureless virtual wall. An all-atom amorphous silica model was also built with surface modification to include hydroxyl groups to provide an intuitive comparison between the virtual and realistic walls. The confinement models and simulation details are described in the next section, followed by results and discussion, and finished with conclusions.

\section{Confinement models and simulation details}

\textbf{Force Fields.} Monte Carlo (MC) simulation is a statistical thermodynamic approach based on the descriptions of fluid-fluid and fluid-pore interactions, i.e., force fields. For the fluid-fluid interaction, the transferable potential for phase equilibria (TraPPE) model \cite{martin1998transferable} was adopted to describe the intra- and inter- molecular interactions. n-Alkanes, such as n-hexane, are nonpolar, flexible chain molecules that are composed of two types of segments, methyl and methylene groups. The TraPPE force field is a united atom model, which means each methyl (-CH$_3$) or methylene (-CH$_2$-) group is considered as a pseudo-atom (i.e., united atom) and carbon atoms are not distinguished from hydrogen atoms when computing the interatomic forces, and therefore abbreviated as TraPPE-UA. The interaction between non-bonded pseudo-atoms (for example the $i^{th}$-atom and $j^{th}$-atom) is calculated from pairwise-additive Lennard-Jones (LJ) 12-6 potential,
\begin{equation}
  U_{\textsf{LJ}}(r_{ij}) = 4\varepsilon_{ij} \left[ \left( \frac{\sigma_{ij}}{r_{ij}}\right)^{12} - \left(\frac{\sigma_{ij}}{r_{ij}} \right)^6\right]
  \label{eqn:LJ-potential}
\end{equation}
in which, $r_{ij}$ is the distance between the $i^{th}$-atom and $j^{th}$-atom; $\varepsilon_{ij}$ and $\sigma_{ij}$ are the well depth and size of Lennard-Jones potential and defined following the Lorentz-Berthelot combining rules, i.e., $\sigma_{\textsf{ij}} = (\sigma_{\textsf{i}} + \sigma_{\textsf{j})}/2$, $\varepsilon_{\textsf{ij}} = \sqrt{\varepsilon_{\textsf{i}}\varepsilon_{\textsf{j}}}$; $\sigma$ and $\varepsilon$ for methyl and methylene groups are listed in \autoref{tbl:LJ_para}. Besides the non-bonded interaction, a set of bonded interactions govern bond stretching, bond-angle bending, and torsional motions. In the TraPPE-UA force field, all bond lengths are fixed as 1.54\si{\angstrom} and thus bond-stretching doesn't contribute to the total potential energy. A harmonic potential is used to control the bond-angle bending, i.e.,
\begin{equation}
  U_{\textsf{bend}}(\theta) = \frac{k_\theta}{2}(\theta - \theta_{\textsf{eq}})^2
  \label{eqn:trappe-bending}
\end{equation}
in which, $\theta$ is the bending angle, $K_{\theta} / k_B = 62500K$ and the equilibrium angle $\theta_{\textsf{eq}} = \ang{114}$. The torsional potentials used to restrict the dihedral rotations around the bonds is based on the force field of optimized potentials for liquid simulations (OPLS-UA) \cite{jorgensen1984optimized}, i.e.,
\begin{equation}
  U_{\textsf{tor}}(\varphi) = c_0 + c_1(1+\textsf{cos}\varphi) +c_2(1-1-\textsf{cos}2\varphi) +c_3(a+\textsf{cos}3\varphi)
  \label{eqn:trappe-torsion}
\end{equation}
in which $\varphi$ is the torsion angle, $c_0/k_B = 0.0$K, $c_1/k_B = 355.03$K, $c_2/k_B=-68.19$K, and $c_3=791.32$K.

For the fluid-pore force field, the interaction between the fluid and slit pore was described by the 10-4-3 Steele potential \cite{steele1973physical, siderius2011extension} and expressed as:
\begin{equation}
  U_{\textsf{steele}}(z) = 2\pi\rho_{\textsf{w}}\sigma_{\textsf{fw}}^2\varepsilon_{\textsf{fw}}\Delta\left[\frac{2}{5}\left(\frac{\sigma_{\textsf{fw}}}{z}\right)^{10} - \left(\frac{\sigma_{\textsf{fw}}}{z}\right)^{4} - \frac{\sigma_{\textsf{fw}}^4}{3\Delta(z+0.61\Delta)^3} \right] \label{eqn:steele-potential}
\end{equation}
The independent variable $z$ in \autoref{eqn:steele-potential} is the relative position of fluid molecule to the pore wall. By using the Steele potential, the pore wall is considered as a structureless surface, which means the pore structure is simplified as a homogeneous virtual surface without atomic details. The Steele force is readily derived from the Steele potential, i.e.,
\begin{equation}
\begin{aligned}
    F_{\textsf{steele}}(z) &= -\partial U_{\textsf{steele}}(z) / \partial z \\
    &=-2\pi\rho_{\textsf{w}}\sigma_{\textsf{fw}}^2\varepsilon_{\textsf{fw}}\Delta
    \left[
    \frac{4}{z}\left(\frac{\sigma_{\textsf{fw}}}{z}\right)^{10} - \frac{4}{z}\left(\frac{\sigma_{\textsf{fw}}}{z}\right)^{4} - 
    \frac{\sigma_{\textsf{fw}}^4}{\Delta(z+0.61\Delta)^4}
    \right]
    \label{eqn:steele-force}
\end{aligned}
\end{equation}
The parameters in \autoref{eqn:steele-potential} and \autoref{eqn:steele-force} are determined according to their physical meanings. $\rho_w$ is the density of the atom constructing the wall; $\Delta$ is the spacing between the upper and lower slab of the slit pore; $\varepsilon_{\textsf{fw}}$ and $\sigma_{\textsf{fw}}$ are the potential depth and size of Lennard-Jones potential and defined following the Lorentz-Berthelot combining rules, i.e., 
$\sigma_{\textsf{fw}} = (\sigma_{\textsf{f}} + \sigma_{\textsf{w})}/2$, $\varepsilon_{\textsf{fw}} = \sqrt{\varepsilon_{\textsf{f}}\varepsilon_{\textsf{w}}}$. Here, we provided two sets of parameters for the walls of single-layered graphene and crystalline quartz, whose $\sigma_{\textsf{w}}$ and $\varepsilon_{\textsf{w}}$ are tabulated in \autoref{tbl:LJ_para} correspondingly. The cutoff distance was chosen as five times of the LJ size, i.e., $r_{\textsf{cutoff}} = 5\sigma_{\textsf{fw}}$.

\begin{table}[ht]
  \caption{Parameters in the Lennard-Jones (\autoref{eqn:LJ-potential}) and Steele potentials (\autoref{eqn:steele-potential})}
  \label{tbl:LJ_para}
  \centering
  \begin{tabular}{cccc}
    \toprule
    & $\varepsilon/k_B(K)$ & $\sigma(\si{\angstrom})$ & $\rho_\textsf{w}(nm^{-3})$  \\
    \midrule
    methyl group (-CH$_3$)   & 98 & 3.75 & -\\
    methylene group (-CH$_2$-) & 46 & 3.95 & -\\
    graphene wall\cite{siderius2011extension} & 28 & 3.40 & 113.7\\
    silica wall\cite{neimark1998pore} & 53 & 3.62 & 26.56\\
    \bottomrule
  \end{tabular}
\end{table}

\textbf{Interface intensity coefficient.} To study the effects of fluid-pore interaction from a homogenization perspective, we introduce the interface intensity coefficient $\lambda$ to the original Steele potential. Then, the Steele potential and force were modified as follows:

\begin{equation}
  U_{\textsf{steele}}^{*}(z) = \textcolor{red} {\lambda} \cdot U_{\textsf{steele}}(z); \quad F_{\textsf{steele}}^*(z) = \textcolor{red} {\lambda} \cdot F_{\textsf{steele}}(z)
  \label{eqn:steele-mod}
\end{equation}

\begin{figure}[ht]
    \centering
    \includegraphics[width=0.8\textwidth]{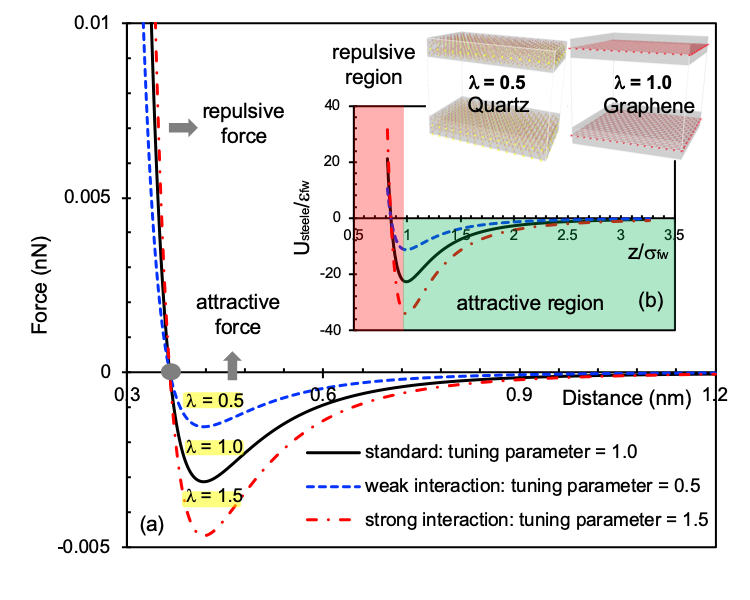}
    \caption{(a) The Steele force and (b) Steele potential with the interface intensity coefficient as a function of the relative distance of hydrocarbon molecular to the pore wall.}
    \label{fig:steele-tune}
\end{figure}

The essential idea of introducing the coefficient $\lambda$ is to enable a systematic study of the effect of varying fluid-pore interactions induced by changes in surface wettability or chemistry. The surface could be modified to be more hydrophilic by adding hydroxyl groups to the pore surface, while the surface could be made more hydrophobic by adding methyl groups. Considering the virtual wall model represented by the Steele potential, the parameters $\varepsilon_\textsf{w}$ and $\rho_\textsf{w}$ are linearly proportional to the potential and force, and thus it is reasonable to extract the overall effects to be represented by the interface intensity coefficient. Although $\sigma_\textsf{w}$ has a complex relationship with the potential and force, it usually does not change too much for the pore materials of our interests. For example, the LJ well position $\sigma$ has only $6\%$ difference between single-layer graphene and crystalline silica systems, as listed in \autoref{tbl:LJ_para}, while the differences from the potential depth and surface density result in a $50\%$ reduction of the fluid-pore interaction of silica wall relative to graphene wall. In this way, we can roughly estimate that the pore-wall interactions of hydrocarbon fluids confined in silica pore is about $50\%$ of that confined in a graphene pore. By setting the graphene wall as a standard reference, the fluid-pore interaction for the silica case is represented by $\lambda = 0.5$. On the other hand, we can also increase the coefficient $\lambda$ (such as 1.5) to create a pore which has a stronger interaction with the fluids relative to the graphene wall. The stronger interaction could be achieved by modifying the surface chemistry of pores with varying sizes of alkyl groups as pointed out by \citet{lowry2018effect}. \autoref{fig:steele-tune} shows the variations of the Steele potential (\autoref{fig:steele-tune}(a)) and force (\autoref{fig:steele-tune}(b)) with the coefficients of 0.5, 1.0 and 1.5 as a function of the relative distance of hydrocarbon molecular to the pore wall. It is seen that the coefficient influences the potential and force more significantly in the attractive region (closer to the point with zero force, shown as a grey dot in \autoref{fig:steele-tune}(a)). When a fluid molecule appears in the repulsive region, the interactions between the molecule and the pore wall are almost the same for $\lambda = 0.5$, 1.0, and 1.5. However, if a fluid molecule moves in to the attractive region, the attractive forces between the molecule and the pore wall have a significant difference for $\lambda = $0.5, 1.0 and, 1.5. This observation reveals that the mediated interaction between the hydrocarbon fluids and the pore wall mainly contributes to the attractive force and excluded volume effects are held approximately constant.

\begin{figure}[ht]
    \centering
    \includegraphics[width=0.8\textwidth]{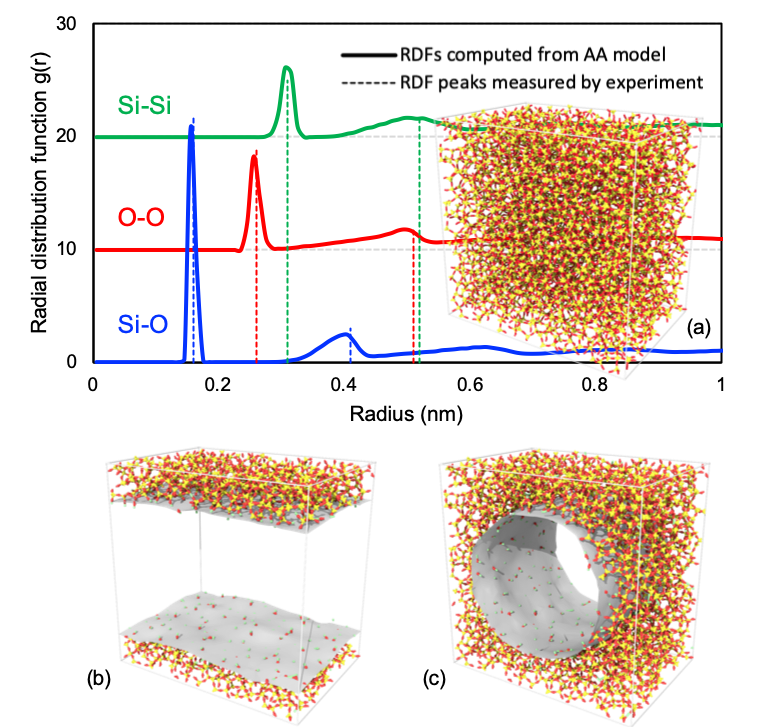}
    \caption{All-atom model of \ce{SiO2}(am): (a) RDFs of \ce{SiO2}(am) and the \ce{SiO2}(am) block; (b) \ce{SiO2}(am) slit pore with $10\%$ -OH coverage; (c) \ce{SiO2}(am) cylindrical pore with $7.2\%$ -OH coverage.}
    \label{fig:RDF}
\end{figure}

\textbf{All-atom model of amorphous silica (\ce{SiO2}(am)).} In addition to the virtual wall described by the Steele potential, we also built a realistic \ce{SiO2}(am) pore structure, as shown in \autoref{fig:RDF}. The purpose is to directly compare the results obtained for the virtual wall and the realistic wall of \ce{SiO2}(am). The interactions of silicon and silicon (Si-Si) atoms, silicon and oxygen (Si-O) atoms, and oxygen and oxygen (O-O) atoms are described by the CLAYFF \cite{cygan2004molecular} force field, and then the all-atom (AA) model of \ce{SiO2}(am) was constructed by a heating and quenching process using a large-scale atomic/molecular massively parallel simulation (LAMMPS) \cite{plimpton1993fast}. 

The radial distribution functions (RDF) of the annealed \ce{SiO2}(am) were plotted for pairs of Si-Si, Si-O, and O-O in \autoref{fig:RDF}. The peaks of RDF computed from the AA model of \ce{SiO2}(am) show a good agreement with those measured by X-ray scattering experiments\cite{mozzi1969structure}. The slit and cylindrical pores are the two pore geometries that are commonly observed in natural shale systems and widely explored in modeling and simulations. In our study, both slit and cylindrical pores were built by removing the atoms in the pore region from the block \ce{SiO2}(am) (\autoref{fig:RDF}(a)), as shown in \autoref{fig:RDF}(b). The volume of the block \ce{SiO2}(am) was 6.2nm*6.2nm*4.2nm (length*width*height). Dangling bonds appear after the deletion of atoms, and hydroxyl groups (-OH) were added to maintain the charge neutrality. For the slit pore, the pore size along the vertical direction was 4nm, and the coverage of -OH group was 15.4/nm$^2$. For the cylindrical pore, the pore radius was 2nm, and the coverage of -OH group was 9.7/nm$^2$.

\begin{figure}[ht]
    \centering
    \includegraphics[width=0.8\textwidth]{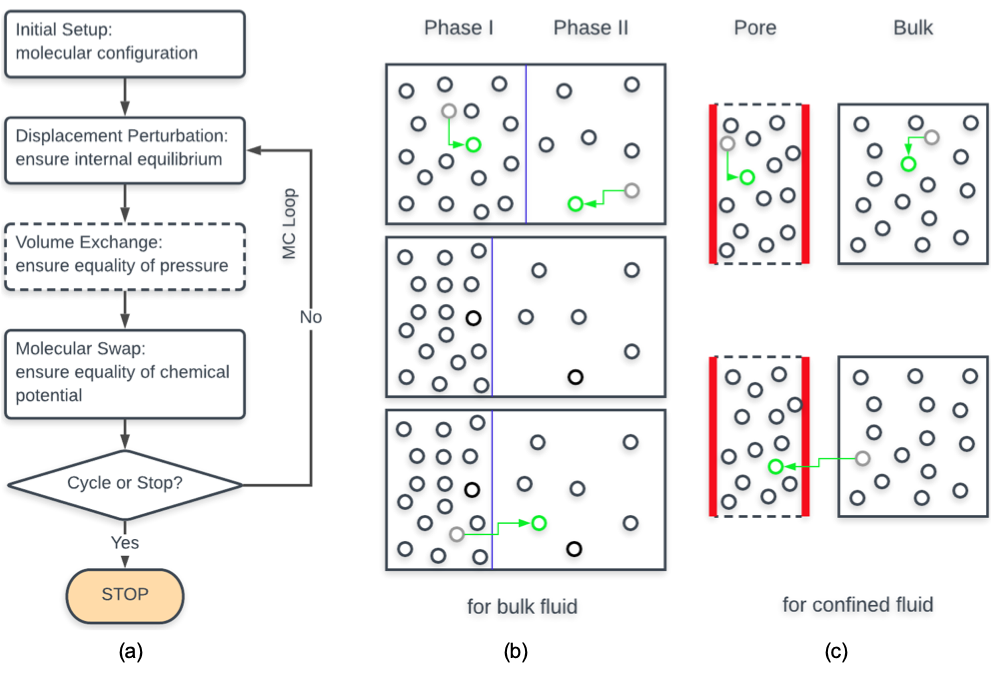}
    \caption{The GEMC method: (a) simulation workflow; (b) for fluids in bulk environment; (c) for confined fluids.}
    \label{fig:GEMC_flowchart}
\end{figure}

\textbf{The nvt-GEMC simulation.} We used the nvt-GEMC method to study the phase equilibrium of hydrocarbon fluids in bulk and confinement environments. During the simulations, the total number of molecules ($N$), phase volumes (V), and the system temperature (T) are kept as constants. The workflow and simulation setup were illustrated in \autoref{fig:GEMC_flowchart}. Generally speaking, there are three types of random walks performed for the fluids in bulk environment. These include displacement perturbation to ensure internal equilibrium, volume exchange to ensure the mechanical equilibrium (i.e., equality of pressure), and molecular swap between boxes to ensure the equality of chemical potential. On the other hand, for the study of confined fluids, the volume exchange was skipped so as to keep the pore size unchanged. The condition for mechanical equilibrium is always automatically satisfied when the chemical potentials of the fluids in the pore and in the bulk are equal as explained by \citet{panagiotopoulos1987adsorption}. 

The numerical investigations were performed using the open-source code of Monte Carlo for Complex Chemical Systems (MCCCS Towhee) \cite{martin2013mcccs}. Following the GEMC workflow, the Monte Carlo simulations started from the displacement perturbation for each simulation box which involved both a center-of-mass molecule translation move and a rotation about the center-of-mass move based on the initial setup of molecular configuration. The volume exchange was conducted for the pair of the two boxes under the constraint that the total volume of the two boxes was kept unchanged. The molecular swap involved configurational-bias two box molecule transfer move and configurational-bias partial molecule regrowth move to increase the acceptance rate, which is especially important for long-chain and cyclic molecules \cite{martin2006using}. The three types of random walk were conducted as a cycle in the MC loop. To obtain statistical averages of physical quantities with reasonable accuracy, the MC cycle was repeated 20,000 times, and the pressure, chemical potential, density, and mole fraction were recorded every 20 cycles and averaged for the last 5,000 cycles for each case discussed in the following section.

\section{Results and Discussion}
\textbf{Bulk systems.} The phase behaviors of hydrocarbons in the bulk environment have been adequately studied by experiments and the theoretical model of the Peng-Robinson Equation of State (PR-EOS). We performed the GEMC simulations for bulk n-hexane (nC$_6$) and compared them with the predictions from the PR-EOS to validate the force field (TraPPE-UA) used for n-hexane and to demonstrate the feasibility of nvt-GEMC to explore the phase behavior of hydrocarbons. One advantage of GEMC for phase equilibrium studies is that it uses two simulation boxes to separate the liquid and vapor phases naturally. When the equilibrium state is attained, the vapor phase is sampled in one of the boxes and the liquid phase in the other, as shown in \autoref{fig:GEMC_flowchart}(b).

\begin{figure}[ht]
    \centering
    \includegraphics[width=0.6\textwidth]{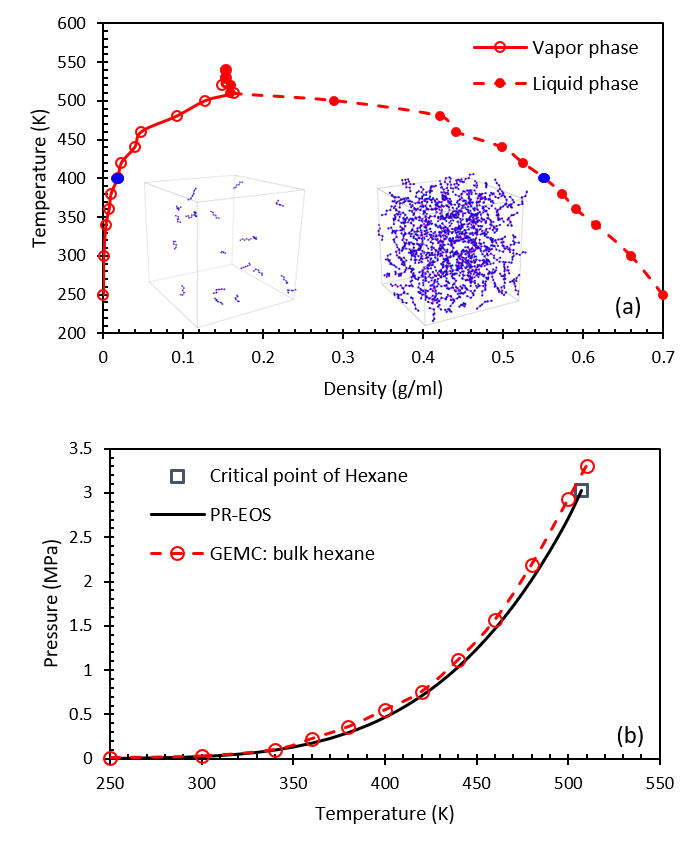}
    \caption{For bulk heptane: (a) variations of densities for liquid and vapor phases as temperature increases; and (b) pressure-temperature diagram.}
    \label{fig:bulk}
\end{figure}

The sizes of the two simulations boxes used in this case were 6nm$*$6nm$*$6nm and 4nm$*$4nm$*$4nm, and the number of n-hexane molecules was chosen as 300. Simulations were performed over a series of temperatures ranging from 250K to 510K, considering that the critical temperature of n-hexane is 507K. For each run, we monitored the pressures and densities of the two simulation boxes. It was worthwhile to emphasize that the equilibrium state at each temperature also established the vapor-liquid coexistence state simultaneously. \autoref{fig:bulk}(a) showed the density variations of each box as the temperature increased. The branch with higher density denoted the liquid phase, and the one with lower density denoted the vapor phase. The snapshots of molecular distributions in the two simulation boxes at the equilibrium states of 400K were inserted in \autoref{fig:bulk}(a). When the density of liquid phase equals that of vapor phase, the critical temperature and pressure of bulk n-hexane were identified as (510K, 3250kPa). Meanwhile, the recorded pressures were plotted against the temperature in \autoref{fig:bulk} (b), in which the red dotted line was obtained from nvt-GEMC simulations and the black solid line was predicted by PR-EOS. It was seen that the results from the GEMC simulation showed a good agreement with the predictions from PR-EOS. Compared with the experimental measurements of critical points (507.44K, 3031kPa), the GEMC simulation overestimated the critical temperature and pressure (less than 7$\%$), yet the errors could be further reduced by performing more case studies for different temperatures near the critical point. Following the same simulation setup for the bulk system, we proceeded to the studies for confined fluids.

\textbf{Van der Waals-type adsorption.} Another advantage of the nvt-GEMC is that it can capture the full phase diagram of confined fluids in the form of a van der Waals loop, which includes stable, metastable, and unstable equilibrium states. The limited capacity of the fluids in the finite simulation box constrain the density fluctuations of fluids in confinement and make it possible to numerically stabilize the fluids in any states, which could be metastable or unstable. Two slit pores types are discussed here, one of which was a graphene-based virtual wall described by the Steele potential and the other one was all-atom \ce{SiO2}(am) with surface modification using -OH groups. To construct the van der Waals loop, isothermal adsorption simulations were performed at 380K for systems with different initial number of molecules ranging from 50 to 400. The condition of vapor-liquid phase coexistence was determined by the Maxwell rule of equal area \cite{neimark2000gauge}. Note that the pore size and morphology (e.g., slit) of the graphene-based virtual wall and all-atom \ce{SiO2}(am) wall were held constant. 

\begin{figure}[ht]
    \centering
    \includegraphics[width=0.6\textwidth]{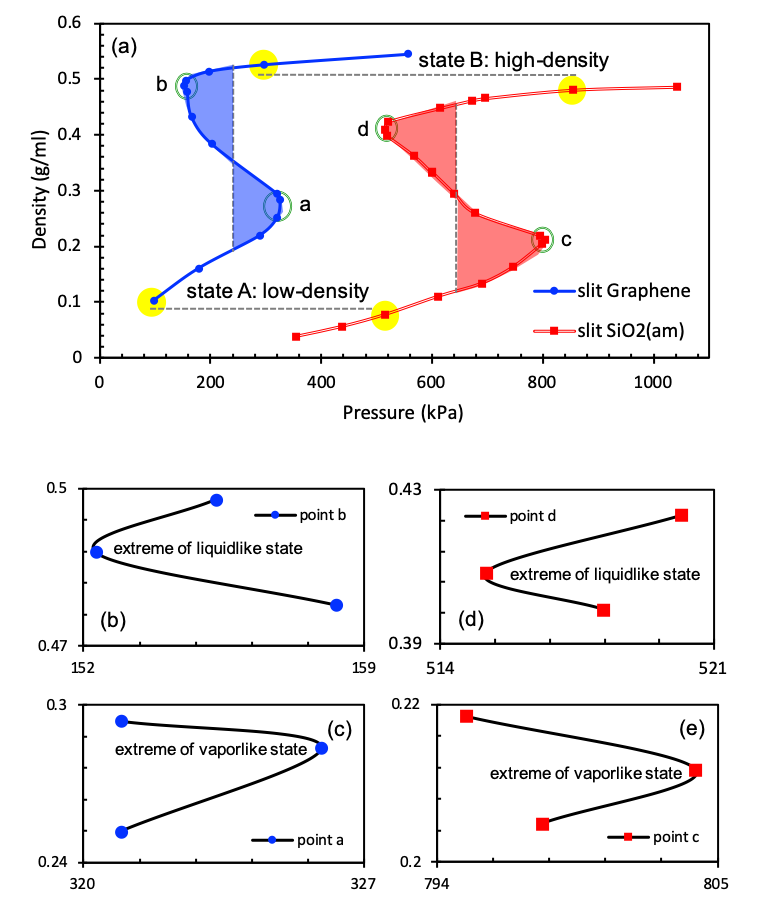}
    \caption{(a) Isothermal adsorptions of confined Heptane under graphene-based virtual wall and all-atom \ce{SiO2}(am) wall at 380K with the two extreme limits enlarged in (b)-(e) correspondingly.}
    \label{fig:van-loop}
\end{figure}

\autoref{fig:van-loop} shows the comparison of isothermal adsorptions of n-hexane at 380K under confinement in the graphene-based virtual wall and in the all-atom \ce{SiO2}(am) wall. The isothermal adsorption of confined n-hexane under the confinement of the graphene-based virtual wall is denoted by the blue single line with circle markers, and that of confined n-hexane in all-atom \ce{SiO2}(am) is denoted by the red double line with square markers. The S-shaped adsorption curve exhibited as a typical van der Waals-type loop with two spinodals, which were the two extreme limits of stability of the vapor-like and liquid-like states, as shown by points a and b or points c and d in \autoref{fig:van-loop}. The compressibility of the confined system was computed from the slope of the van der Waals loop. It was observed that the slopes of vapor branches (from low-density states to the limits of vapor-like states, points a and c) were larger than the slopes of liquid branches (from the limits of liquid-like states, points b and d, to high-density states). Also, the slopes of the unstable branches (the backward trajectories from point a to b or from point c to d) were negative. The results were consistent with the physical understanding that the compressibilities of fluids in vapor phase are usually larger than the liquid phase, and the negative compressibilities corresponded to unstable states. More importantly, the results showed that the compressibility of confined fluids was highly influenced by the strength of fluid-wall interactions based on the observation that the compressibility of n-hexane confined in graphene pore was larger than that in \ce{SiO2}(am) pore.

\begin{figure}[ht]
    \centering
    \includegraphics[width=0.6\textwidth]{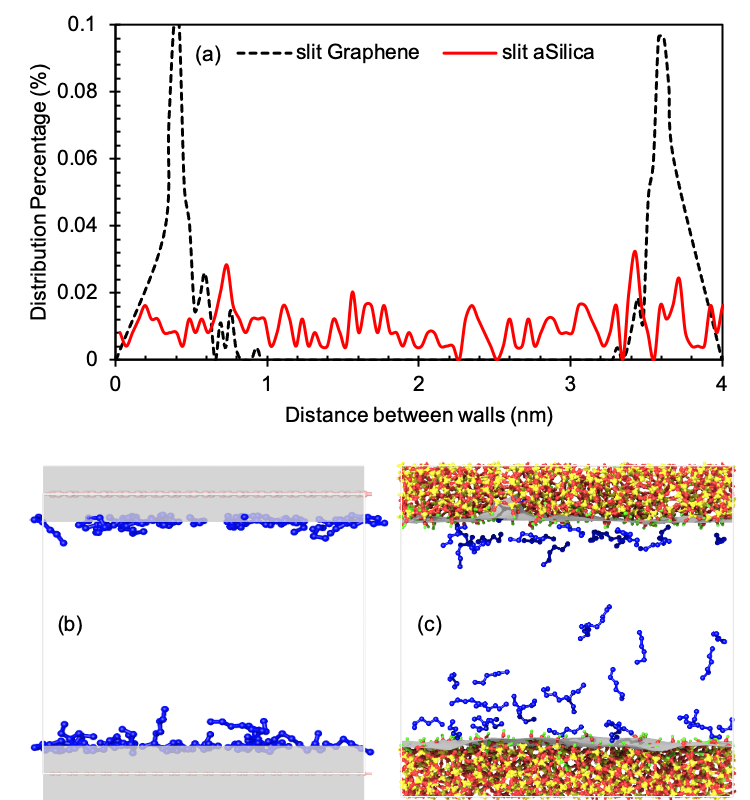}
    \caption{Molecular distributions of state A for graphene and \ce{SiO2}(am) pores: (a) the comparison of distribution percentage across the pore walls; (b) the molecular distribution in the graphene slit pore; and (c) the molecular distribution in the \ce{SiO2}(am) pore.}
    \label{fig:low-distribution}
\end{figure}

To understand how confinement mediated the phase behavior of hydrocarbon fluids, we further compared the molecular distribution across the pore wall for both of the two above-mentioned cases. Two representative states, state A with a low fluid density and state B with a high fluid density (highlighted in \autoref{fig:van-loop}), were explored in detail. For state A of low fluid density, \autoref{fig:low-distribution} showed the molecular distribution of n-hexane confined in slit graphene and all-atom \ce{SiO2}(am) pores. \autoref{fig:low-distribution}(a) provided a direct comparison of the molecular distribution across the two types of pore walls. The black dotted line denotes the molecular distribution across the graphene wall, and the red solid line denotes the molecular distribution across the \ce{SiO2}(am) wall. The distribution percentage was computed as the number of n-hexane molecules at different locations divided by the total number of n-hexane molecules in the confined system with the resolution of 100 bins across the wall. As indicated by the two peaks of the black dotted line, most n-hexane molecules were adsorbed to the surface of the graphene pore wall, while only part of the n-hexane molecules was adsorbed to the surface of the \ce{SiO2}(am) pore wall. The snapshots of the two systems at equilibrium states were depicted in \autoref{fig:low-distribution}(b) and (c). It was readily concluded that the interactions between the n-hexane and graphene-based virtual walls were stronger than the interactions between the n-hexane and all-atom \ce{SiO2}(am) wall with -OH surface modifications. 

\begin{figure}[ht]
    \centering
    \includegraphics[width=0.6\textwidth]{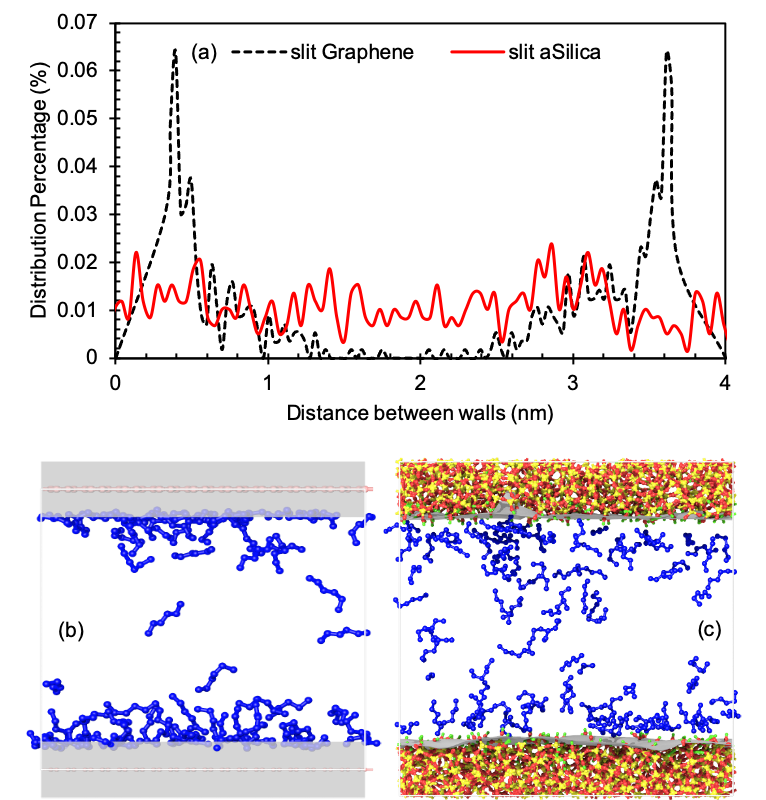}
    \caption{Molecular distributions of state B for graphene and \ce{SiO2}(am) pores: (a) the comparison of distribution percentage across the pore walls; (b) the molecular distribution in the graphene slit pore; and (c) the molecular distribution in the \ce{SiO2}(am) pore.}
    \label{fig:high-distribution}
\end{figure}

For state B of high fluid density, the molecular distribution across the wall is shown in \autoref{fig:high-distribution}. The black dotted line also exhibited two peaks close to the walls. Different from state A, some n-hexane molecules of state B appeared in the middle of graphene-based virtual walls because the n-hexane had fully saturated near the pore surface. The molecular distribution of n-hexane in the all-atom \ce{SiO2}(am) wall, denoted by the red solid line in \autoref{fig:high-distribution}, showed similar distribution features as state A, and confirmed the observation from state A that the attraction of all-atom \ce{SiO2}(am) wall with -OH surface modification was very weak compared with the graphene-based virtual wall. From the comparison of the van der Waals loops obtained for the two types of pore walls, it was concluded that the attraction force between hydrocarbons and the graphene-based virtual wall was stronger than the all-atom \ce{SiO2}(am) wall with -OH surface modification, and resulted in a lower saturated vapor pressure as shown in \autoref{fig:van-loop}. This direct comparison demonstrated the importance of the fluid-pore interaction in mediating the phase behaviors of hydrocarbon fluids.

\begin{figure}[ht]
    \centering
    \includegraphics[width=0.6\textwidth]{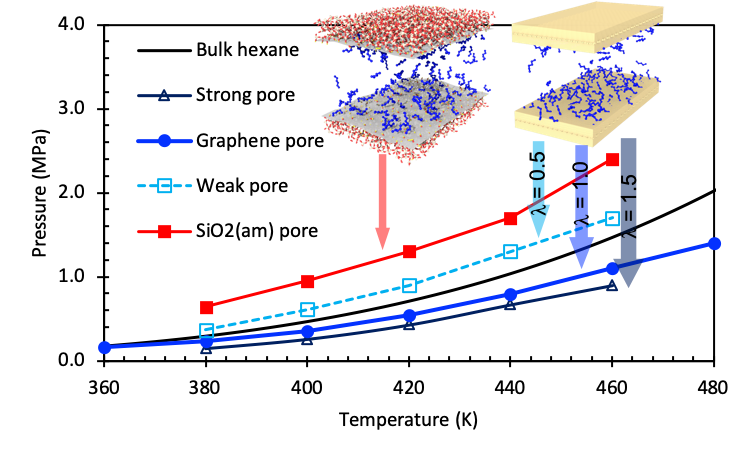}
    \caption{Pressure temperature diagrams of n-hexane under the confinements of strong pore, graphene pore, weak pore and \ce{SiO2}(am) pore.}
    \label{fig:wet-diagram}
\end{figure}

\textbf{Fluid-pore interaction effect.} In this section, we introduced more types of surfaces to understand the effect of surface energy (i.e., the strength of fluid-pore interaction) on the saturation pressure of confined fluids. For the all-atom \ce{SiO2}(am) wall, the surface energy, wettability and the fluid-pore interaction had to be modified by changing the functional groups attached to the surface area, and thus, the surface properties were difficult to be controlled with the desired surface energy. In contrast, the introduced coefficient $\lambda$ in Eq.~\ref{eqn:steele-potential} provided a straightforward way to tune the surface energy. Taking the case of graphene wall represented by $\lambda = 1.0$ as a reference case, we further studied another two typical cases of $\lambda = 0.5$ and $\lambda = 1.5$ which described surfaces with weaker and stronger attractions to hydrocarbons compared with the graphene wall. The saturated pressures of confined n-hexane were computed from the van der Waals loops as discussed earlier. \autoref{tab:pressure-temp} listed the pressures (in units of MPa) of extreme points in the van der Waals loops, including equilibrium pressure (EP), the extreme limits of liquid-like state (LP) and vapor-like state (VP) at temperatures from 380K to 460K. The sets of equilibrium pressures was also plotted in \autoref{fig:wet-diagram} to construct the pressure-temperature phase diagrams of confined n-hexane. We had seen that a stronger surface energy led to the lower saturation pressure at 380K by comparing the van der Waals loops of n-hexane confined in the graphene pore and \ce{SiO2}(am) pore with surface modifications of -OH groups. \autoref{fig:wet-diagram} demonstrated that this conclusion was also applied for higher temperatures up to 460K. By varying the coefficient $\lambda$ from 0.5 to 1.0 and 1.5, a similar trend was observed and further confirmed that the higher surface energy (simulated with larger $\lambda$) resulted in the lower saturation pressure. More interestingly, the saturation pressure could be elevated or suppressed relative to the bulk phase as illustrated in \autoref{fig:wet-diagram}. As the surface energy (i.e., fluid-pore interaction) decreased (smaller $\lambda$), the isothermal vapor pressure increased, indicating a greater preference for the fluid to exist in the vapor state. Sufficient reduction of the fluid-pore interactions could even elevate the vapor pressure above that of the bulk fluid. Therefore, it was noted that the pressure-temperature phase diagram of the bulk phase was not the extreme limit of confined systems. Instead, the vapor pressure and temperature of confined systems could be elevated or suppressed depending on the strength of fluid-pore interactions provided by the different types of pore materials and surface treatments.

\begin{table}[ht]
\caption{Equilibrium pressures (EP), extremes of liquid-like state (LP) and vapor-like state (VP) in the unit of MPa for bulk n-hexane and confined n-hexane.}
\begin{center}
\begin{tabular}{cccccccccccccc}
    \hline
    \multicolumn{1}{c}{\multirow{2}{*}{T(K)}}&Bulk&\multicolumn{3}{c}{\ce{SiO2}(am)}&\multicolumn{3}{c}{$\lambda$=0.5}&\multicolumn{3}{c}{$\lambda$=1.0}&\multicolumn{3}{c}{$\lambda$=1.5}\\
    \multicolumn{1}{c}{}&(EP)&(EP&LP&VP)&(EP&LP&VP)&(EP&LP&VP)&(EP&LP&VP)\\
    \hline
    380&0.29 &0.64&0.52&0.81 &0.37&0.21&0.65 &0.23&0.15&0.33 &0.15&0.12&0.21\\
    400&0.46 &0.95&0.88&1.07 &0.61&0.45&0.86 &0.35&0.26&0.43 &0.26&0.20&0.33\\
    420&0.71 &1.31&1.25&1.42 &0.89&0.68&1.15 &0.54&0.45&0.61 &0.43&0.41&0.45\\
    440&1.03 &1.69&1.64&1.72 &1.30&1.17&1.42 &0.79&0.75&0.81 &0.67&0.65&0.69\\
    460&1.46 &2.40&2.34&2.42 &1.71&1.64&1.75 &1.11&1.08&1.16 &0.89&0.88&0.91\\
    \hline
\end{tabular}
\end{center}
\label{tab:pressure-temp}
\end{table}

\autoref{tab:difference} lists the pressure gradients between the extremes of vapor-like states and that of liquid-like states (denoted by VP-LP) for all types of pore walls studied in this work at temperatures ranging from 380K to 460K. Meanwhile, the pressure differences of the extremes of vapor-like states and liquid-like states relative to the saturation pressures were listed in the columns of LPD and VPD. Based on the pressure gradients, the width of the van der Waals loops as well as the levels of the extremes of vapor-like states and liquid-like states relative to the equilibrium states could be examined. As the temperature increased, the pressure gradients calculated from VP-LP, LPD and VPD decreased and the width of the van der Waals loop became narrow. This phenomena satisfied the understanding that when the fluid approached the critical temperature, the property differences between liquid and vapor became less.

\begin{table}[ht]
\caption{The pressure gradients between VP and LP, the pressure gradients between LP and EP (LPD), and the pressure gradients between VP and EP (VPD) of confiend n-hexane in the unit of MPa. }
\begin{center}
\begin{tabular}{ccccccccccccc}
    \hline
    \multicolumn{1}{c}{\multirow{2}{*}{T(K)}}&\multicolumn{3}{c}{\ce{SiO2}(am)}&\multicolumn{3}{c}{$\lambda$=0.5}&\multicolumn{3}{c}{$\lambda$=1.0}&\multicolumn{3}{c}{$\lambda$=1.5}\\
    \multicolumn{1}{c}{}&(VP-LP&LPD&VPD)&(VP-LP&LPD&VPD)&(VP-LP&LPD&VPD)&(VP-LP&LPD&VPD)\\
    \hline
    380 &0.29&-0.12&0.17  &0.44&-0.16&0.28  &0.18&-0.08&0.10  &0.09&-0.03&0.06\\
    400 &0.19&-0.07&0.12  &0.41&-0.16&0.25  &0.17&-0.09&0.08  &0.13&-0.06&0.07\\
    420 &0.17&-0.06&0.11  &0.47&-0.21&0.26  &0.16&-0.09&0.07  &0.04&-0.02&0.02\\
    440 &0.08&-0.05&0.03  &0.25&-0.13&0.12  &0.06&-0.04&0.02  &0.04&-0.02&0.02\\
    460 &0.08&-0.06&0.02  &0.11&-0.07&0.04  &0.08&-0.03&0.05  &0.03&-0.01&0.02\\
    \hline
\end{tabular}
\end{center}
\label{tab:difference}
\end{table}

\begin{figure}[ht]
    \centering
    \includegraphics[width=0.6\textwidth]{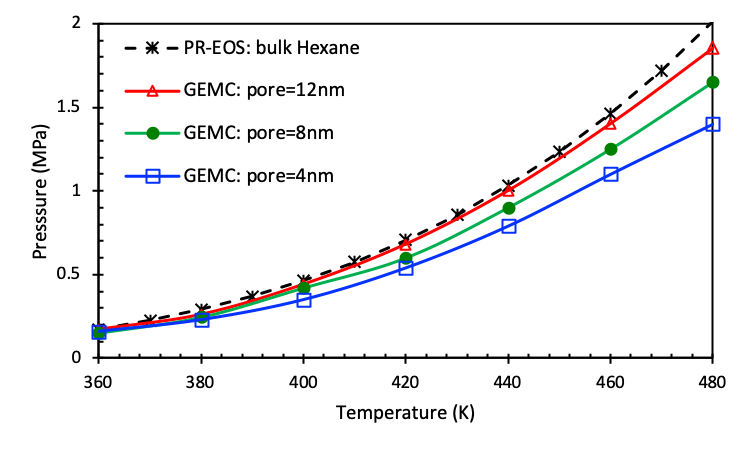}
    \caption{Pressure temperature diagrams of confined n-hexane in graphene-based virtual slit pores of pore sizes of 4nm, 8nm, and 12nm.}
    \label{fig:size}
\end{figure}

\textbf{Pore size and morphology effects.} In addition to the effect of fluid-pore interaction strength, the pore size and morphology also play important roles in altering the phase behaviors of hydrocarbons. \autoref{fig:size} presented the pressure-temperature phase diagrams of n-hexane confined in the graphene-based virtual slit pores of pore sizes of 4nm, 8nm, and 12nm. The pore size was controlled through adjusting the distance between the top and bottom surfaces of the slit pores. The transitional resolution of pore size under studies was 4nm, and the resolution of temperatures was every 20K from 360K to 480K. The plot demonstrated the deviations of pressure-temperature phase diagrams of confined fluids from that in bulk environment for sub- 10nm pores. The deviations from bulk behavior became significant under high temperature and pressure conditions for the confined system at a fixed pore size. Furthermore, it was shown that the phase behaviors of confined fluids approached bulk behavior as the pore size increased to 10nm, and supports previous observations that the effects of pores of size under sub- 10nm disappeared as widely reported in the community of petroleum engineering.

Two commonly observed pore shapes in natural shale systems, and are thus widely modeled. are slit and cylindrical pores. To demonstrate the influence of pore morphology on phase transitions, the flat slit wall and the curved cylindrical wall with same surface areas yet different volumes were constructed and shown in \autoref{fig:RDF}. The curvature was computed as the reciprocal of radius and calculated as 0.5 and 0.0 for the cylindrical surface (the radius is 2nm) and the slit surface (consider the flat surface as a circle with infinite large radius) respectively. The surface curvature contributed to the capillary condensation near the pore surface and resulted in reduced vapor pressure of confined fluids at all temperature levels. However, the quantitative relationships between the surface curvature and pressure enhancement are still unclear and warrant further investigation \cite{srivastava2017pressure}. More discussion on the pore shape and surface curvature effects is planned for future work.

\begin{figure}[ht]
    \centering
    \includegraphics[width=0.6\textwidth]{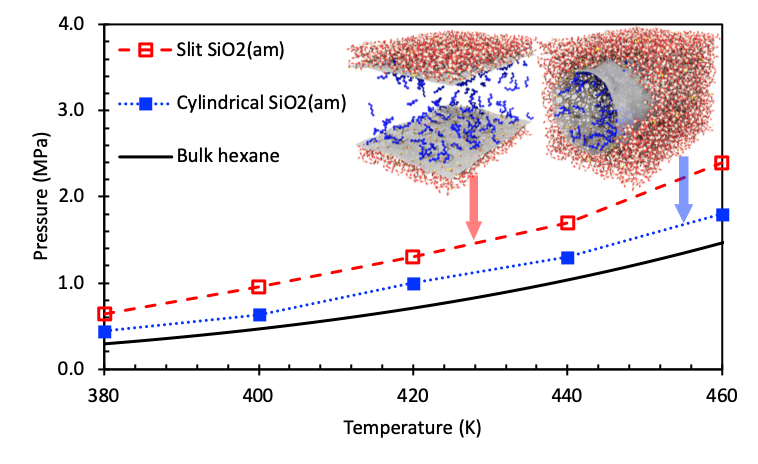}
    \caption{The comparison of pressure temperature diagrams of n-hexane confined in slit \ce{SiO2}(am) pore and cylindrical \ce{SiO2}(am) pore.}
    \label{fig:morphology}
\end{figure}

\section{Conclusion}

In this work, we systematically studied pore effects on the pressure-volume-temperature (PVT) behaviors of confined n-hexane. Three crucial parameters were identified and thoroughly investigated, including the strength of fluid-pore interaction, the pore size, and the pore morphology. The study on the strength of fluid-pore interaction effects was reported for the first time and was demonstrated to play a significant role in mediating the phase behavior of confined fluids. As the surface energy (i.e., fluid-pore interaction) decreased, the vapor pressure increased while the vapor temperature decreased. More importantly, sufficient reduction of the fluid-pore interactions could even elevate the vapor pressure above that of the bulk fluid. The pore size effect on the pressure-temperature diagram was also presented with high transitional resolutions. Results show that when the pore size was greater than 10nm, the phase behavior of confined fluids approached that of the bulk phase, and the effect of confinement becomes less significant in altering the fluid thermodynamic properties. Moreover, the pore morphology effect was studied by directly comparing two shapes of confinement and it was concluded that the surface curvature enhanced the adsorption of hydrocarbons and resulted in a decreased saturation pressure. The conclusions provide a better understanding on the pore effects on the phase behaviors of confined hydrocarbons, explained the inconsistent measurements regarding the elevated or suppressed vapor temperatures by DSC experiments, and ultimately help improve hydraulic fracturing protocols to extract energy from unconventional tight reservoirs.

\section*{Acknowledgement}
This work was supported by EFRC-MUSE, an Energy Frontier Research Center funded by the U.S. Department of Energy, Office of Science, Basic Energy Sciences under Award No. DE-SC0019285. Simulations were performed using resources in the High-Performance Computing Center at Idaho National Laboratory, which is supported by the Office of Nuclear Energy of the U.S. Department of Energy and the Nuclear Science User Facilities under Contract No. DE-AC07-05ID14517.

\clearpage
\bibliographystyle{unsrtnat}

\section*{Graphical Abstract}
\begin{figure}[ht]
    \centering
    \includegraphics[width=3.3in]{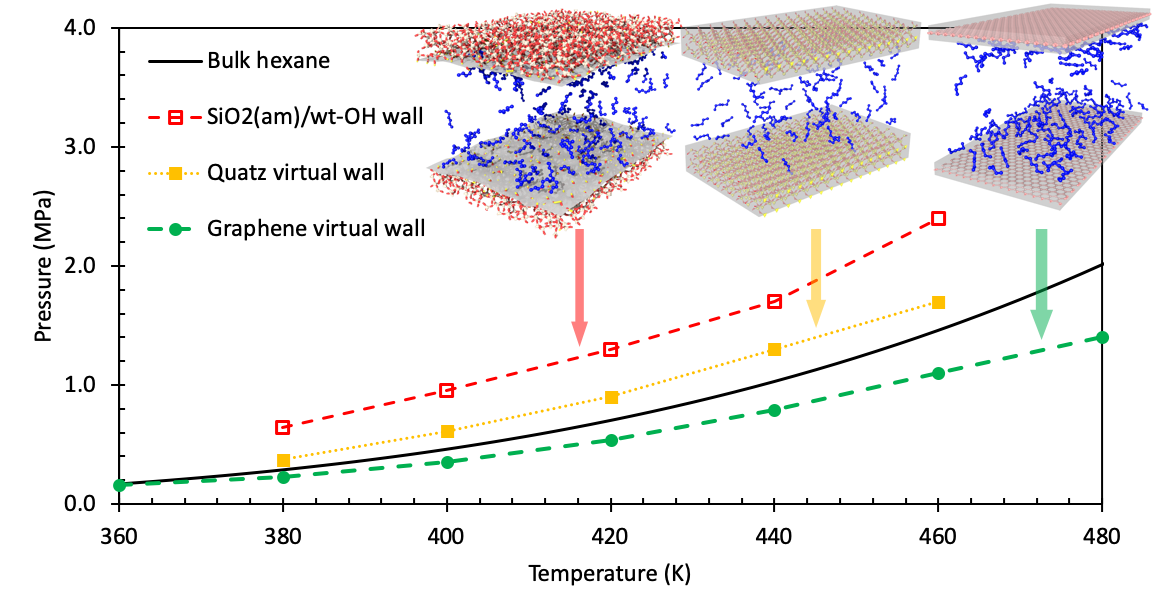}
\end{figure}

\end{document}